\documentclass[a4paper,final]{appolb}
\usepackage{graphicx,amsmath,amssymb,hyperref,lmodern}
\usepackage[numbers]{natbib}

\newcommand{\ud}{\mathrm{d}}
\newcommand{\ve}{\varepsilon}

\begin{document}
\title{Towards generic cosmological evolution without singularity 
and fundamental symmetry of space-time\thanks{Presented at the 3rd Conference of the Polish Society on Relativity, Krak{\'o}w, Poland, September 25--29, 2016.}
}
\author{Orest Hrycyna
\address{Theoretical Physics Division, 
National Centre for Nuclear Research,\\  Ho{\.z}a 69, 00-681 Warszawa, Poland\\ \tt{orest.hrycyna@ncbj.gov.pl}}
}
\maketitle

\begin{abstract}
\noindent
Dynamical systems methods are used to investigate cosmological model with non-minimally coupled scalar field. Existence of an asymptotically unstable de Sitter state distinguishes values of the non-minimal coupling constant parameter $\frac{3}{16}\le\xi<\frac{1}{4}$, which correspond to conformal coupling in higher dimensional theories of gravity. 
\end{abstract}

\section{Introduction}
We begin with action integral for the theory where a gravitational part is given by the Einstein-Hilbert term and a substance filling the universe is described by the non-minimally coupled scalar field 
\begin{equation}
\begin{split}
S&=S_{g}+ S_{\phi} = \\ & \frac{1}{2\kappa^{2}}\int\ud^{4}x\sqrt{-g}\,R - \frac{1}{2}\int\ud^{4}x\sqrt{-g}\Big(\ve\nabla^{\alpha}\phi\,\nabla_{\alpha}\phi + \ve\xi R \phi^{2} + 2U(\phi)\Big)\,,
\end{split}
\end{equation}
where $\kappa^{2}=8\pi G$, $\ve=\pm1$ corresponds to a canonical and a phantom scalar field, respectively, and $\xi$ is dimensionless non-minimal coupling constant between the scalar field and the gravity.

Working with a spatially flat Friedmann-Robertson-Walker metric,
we obtain the energy conservation condition 
\begin{equation}
\label{eq:constr}
\frac{3}{\kappa^{2}}H^{2}=\rho_{\phi}=\ve\frac{1}{2}\dot{\phi}^{2}+U(\phi)+\ve3\xi H^{2}\phi^{2}+\ve6\xi H \phi\dot{\phi}\,,
\end{equation}
the acceleration equation
\begin{equation}
\label{eq:accel}
\dot{H}=-2H^{2}+\frac{\kappa^{2}}{6}\frac{-\ve(1-6\xi)\dot{\phi}^{2}+4U(\phi)-6\xi\phi U'(\phi)}{1-\ve\xi(1-6\xi)\kappa^{2}\phi^{2}}\,,
\end{equation}
and the equation of motion for the scalar field
\begin{equation}
\label{eq:scfield}
\ddot{\phi}+3H\dot{\phi}+6\xi\big(\dot{H}+2H^{2}\big)\phi + \ve U'(\phi)=0\,.
\end{equation}
For given form of the scalar field potential function $U(\phi)$ the system of equations \eqref{eq:accel} and \eqref{eq:scfield} subject to the energy conservation condition \eqref{eq:constr} constitutes dynamical system in variables $(\phi,\dot{\phi},H)$ which completely describes evolution of the model under considerations.

\section{Dynamical system and instability of the initial de Sitter state}

Let us assume that for large values of the scalar field starting form some value $\phi>m_{*}$ the potential function can be approximated as $U(\phi)=\pm M_{1}^{4-m}\phi^{m}\pm M_{2}^{4+n}\phi^{-n}$ with $m+n>0$ where $M_{i}$ are constants. Clearly, the first term dominates while the second term constitutes some small deviation. Next, introducing new dimensionless dynamical variables $u=\frac{\dot{\phi}}{H\phi}=\frac{\ud\ln{\phi}}{\ud\ln{a}}$ and $v=\frac{\sqrt{6}}{\kappa\phi}$, together with dimensionless constants $\alpha_{1}=\pm2\frac{M_{1}^{4-m}}{H_{0}^{2}}\left(\frac{\sqrt{6}}{\kappa}\right)^{m-2}$, $\alpha_{2}=\pm2\frac{M_{2}^{4+n}}{H_{0}^{2}}\left(\frac{\sqrt{6}}{\kappa}\right)^{-n-2}$ we can write the energy conservation condition
\begin{equation}
\frac{H^{2}}{H_{0}^{2}}=\frac{v^{2-m}\big(\alpha_{1}+\alpha_{2}v^{n+m}\big)}{v^{2}-\ve(1-6\xi)u^{2}-\ve6\xi(u+1)^{2}}\,,
\end{equation}
and the acceleration equation
\begin{equation}
\label{eq:accel_2}
\frac{\dot{H}}{H^{2}} = -2 +\frac{-\ve(1-6\xi)u^{2}+\frac{H_{0}^{2}}{H^{2}}v^{2-m}\big(\alpha_{1}(2-3\xi m)+\alpha_{2}(2+3\xi n)v^{n+m}\big)}{v^{2}-\ve6\xi(1-6\xi)}\,.
\end{equation}
Finally, we obtain the following two dimensional dynamical system 
\begin{equation}
\label{eq:dynsys}
\begin{split}
\frac{\ud u}{\ud\ln{a}} & = -u(u+1)-(u+6\xi)\left(\frac{\dot{H}}{H^{2}}+2\right)-\ve\frac{1}{2}\frac{H_{0}^{2}}{H^{2}}v^{2-m}\big(m \alpha_{1}-n\alpha_{2}v^{n+m}\big)\,,\\
\frac{\ud v}{\ud\ln{a}} & = - u v\,,
\end{split}
\end{equation}
where a time parameter along phase space curves is the natural logarithm of the scale factor.

\begin{figure}
\centering
\includegraphics[scale=0.35]{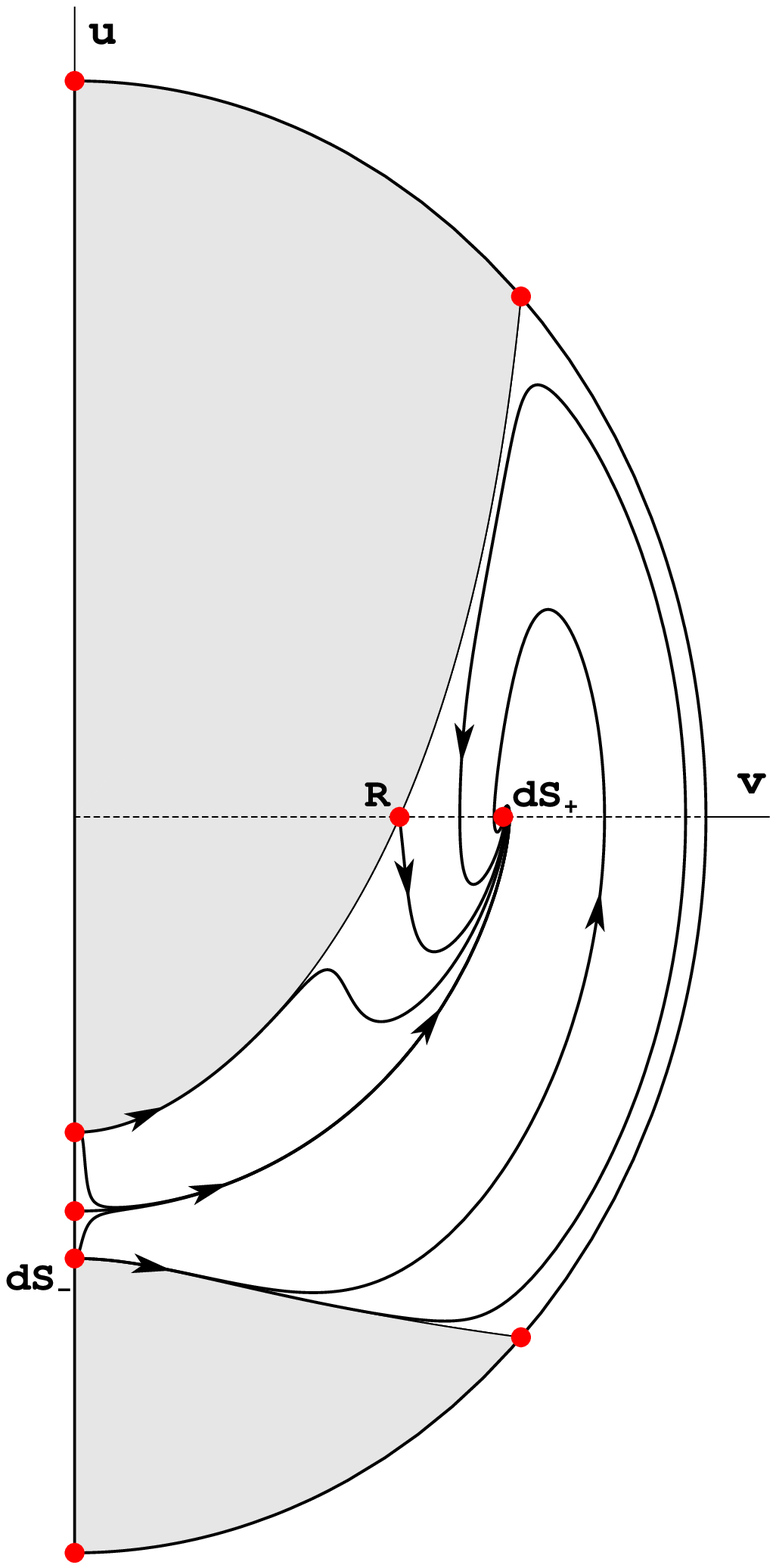}
\hspace{1.5cm}
\includegraphics[scale=0.35]{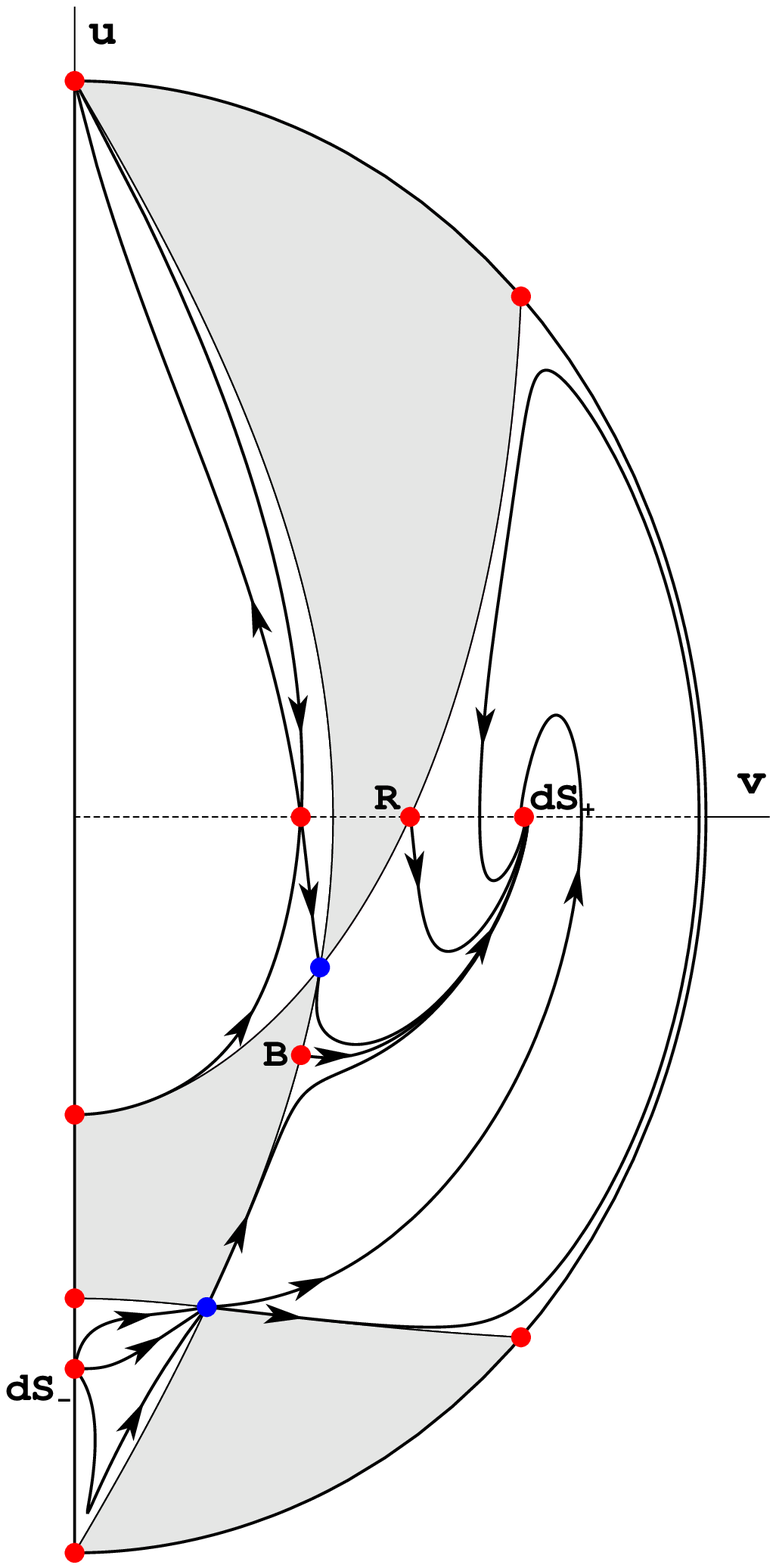}
\caption{Phase space portraits for the canonical scalar field $\ve=+1$ and: $\xi=\frac{3}{16}$, $m=-1$, $n=2$, $\alpha_{1}=1$, $\alpha_{2}=\frac{1}{4}$ (left); $\xi=\frac{3}{14}$, $m=2$, $n=1$, $\alpha_{1}=-1$, $\alpha_{2}=3$ (right). The shaded regions where $H^{2}<0$ are nonphysical. For both phase space diagrams there is an open and dense set of initial conditions leading to the singularity free evolutional paths connecting the unstable during expansion de Sitter state $dS_{-}$ with the stable one $dS_{+}$.}
\vspace{-5mm}
\label{fig:1}
\end{figure}

In what follows, we are interested only in asymptotic states, the critical points, located at infinite values of the scalar scalar field, i.e.~for which $v^{*}=0$. Additionally we are looking for states where the acceleration equation \eqref{eq:accel_2} vanishes. One can easy find that this situation takes place in two cases: first for $m=2$, which corresponds to an asymptotically quadratic potential function, we have the phase space coordinate $u^{*}=-\frac{2\xi}{1-4\xi}$, second for $\xi=\frac{3}{16}$ we have $u^{*}=-\frac{3}{2}$. Using eigenvalues of the linearisation matrix in the vicinity of those critical points one can obtain stability conditions for corresponding de Sitter state which give rise to additional constraints on values of the non-minimal coupling constant in the first case and on the asymptotic form of the potential function in the second case. In figure \ref{fig:1} we present phase space diagrams 
for two cases under considerations.  

Let us discuss the case with an asymptotically quadratic scalar field potential function with $m=2$. The eigenvalues of the linearisation matrix of the system \eqref{eq:dynsys} in the vicinity of the critical point $u^{*}=-\frac{2\xi}{1-4\xi}$, $v^{*}=0$ are $\lambda_{1}=-4+\frac{1}{1-4\xi}$, $\lambda_{2}=\frac{2\xi}{1-4\xi}$. It is clear that both eigenvalues are positive when $\frac{3}{16}<\xi<\frac{1}{4}$ which give rise to an unstable during expansion of the universe critical point. Using linearised solutions in the vicinity of this state one can calculate value of the Hubble function \eqref{eq:constr} at the initial de Sitter state which is
\begin{equation}
\left(\frac{H^{*}}{H_{0}}\right)^{2}=-\ve\alpha_{1}\frac{(1-4\xi)^{2}}{2\xi(1-6\xi)(3-16\xi)}\,,
\end{equation} 
where $\ve\alpha_{1}<0$. Taking the canonical scalar field $\ve=+1$ and $\alpha_{1}=-2\frac{M_{1}^{2}}{H_{0}^{2}}$ for $m=2$ we find that the energy density at this state can be smaller than the Planck energy density
\begin{equation}
2\frac{M_{1}^{2}}{m_{Pl}^{2}}\frac{(1-4\xi)^{2}}{2\xi(1-6\xi)(3-16\xi)}<\frac{8\pi}{3}\,,
\end{equation}
even for mass of the scalar field $2 M_{1}^{2}\simeq m_{Pl}^{2}$.

\section{Conclusions}

In the most simple case of minimally coupled scalar field the equation of motion is conformally invariant only in $2$ space-time dimensions. Due to presence of the non-minimal coupling term this property can be generalised to $d>2$ space-time dimensions. Assuming a monomial potential function the Klein-Gordon equation for the scalar field
$
\square\phi-\xi R\phi-\ve \alpha M^{4-\alpha}\phi^{\alpha-1}=0\,,
$
is conformally invariant only if
$
\xi=\xi_{\textrm{conf}}=\frac{1}{4}\frac{d-2}{d-1}\,,\quad \alpha=\alpha_{\textrm{conf}}=\frac{2d}{d-2}\,.
$
In this way we obtain a discrete set of theoretically allowed values of the non-minimal coupling constant suggested by the conformal invariance condition of the scalar field in $d\ge2$ space-time dimensions
$
\big\{(d,\xi,\alpha)\big\}=\Big\{\left(2,0,\infty\right),\left(3,\frac{1}{8},6\right),\left(4,\frac{1}{6},4\right), \left(5,\frac{3}{16},\frac{10}{3}\right),\dots,\left(\infty,\frac{1}{4},2\right)\Big\}\,.
$

In this short note we have used dynamical systems methods to investigate FRW cosmological model with non-minimally coupled scalar field and an asymptotically monomial potential function. Performed analysis and existence of the de Sitter state which is unstable during expansion of the universe give firm constraints on the non-minimal coupling constant. From the other side obtained values of the non-minimal coupling constant $\frac{3}{16}\le\xi<\frac{1}{4}$ correspond to the values suggested by the conformal coupling condition in higher dimensional theories of gravity. This suggest that the conformal invariance should be considered as a serious candidate for the fundamental symmetry of the space-time not only in the substantial sector of the theory but also in the gravitational \cite{tHooft:2014daa}. Additionally observational cosmological constraints on the non-minimal coupling constant suggest similar values \cite{Hrycyna:2015vvs}.

\noindent
\\
This research was supported by the National Science Centre through the OPUS 5 funding scheme (Decision No.~DEC-2013/09/B/ST2/03455).

\providecommand{\href}[2]{#2}\begingroup\raggedright\endgroup

\end{document}